\documentclass[aps,twocolumn,showpacs,amsmath,amssymb,prb]{revtex4}

\usepackage[dvips]{graphicx}
\usepackage{dcolumn}
\usepackage{bm}

\begin{document}


\title{Pressure-induced Phonon Softenings and the
Structural and Magnetic Transitions in CrO$_{2}$}

\author{Sooran Kim, Kyoo Kim, Chang-Jong Kang, and B. I. Min*}
\affiliation{Department of physics, PCTP,
Pohang University of Science and Technology, Pohang, 790-784, Korea}

\begin{abstract}
To investigate the pressure-induced structural transitions of
chromium dioxide (CrO$_{2}$),
phonon dispersions and total energy band structures
are calculated as a function of pressure.
The first structural transition has been confirmed at P$\approx$ 10 GPa
from the ground state tetragonal CrO$_{2}$ (t-CrO$_{2}$) of rutile type
to orthorhombic CrO$_{2}$ (o-CrO$_{2}$) of CaCl$_{2}$ type.
The half-metallic property is found to be
preserved in o-CrO$_{2}$. The softening of Raman-active B$_{1g}$ phonon mode,
which is responsible for this structural transition, is demonstrated.
The second structural transition is found to occur for P$\geq$ 61.1 GPa
from ferromagnetic (FM) o-CrO$_{2}$ to nonmagnetic (NM)
monoclinic CrO$_{2}$ (m-CrO$_{2}$) of MoO$_{2}$ type,
which is related to the softening mode
at {\bf q} = R($\frac{1}{2}$,0,$\frac{1}{2}$).
The third structural transition has been newly identified at P= 88.8 GPa
from m-CrO$_{2}$ to cubic CrO$_{2}$ of CaF$_{2}$ type
that is a FM insulator.

\end{abstract}


\maketitle

\section{Introduction}
 Chromium dioxide (CrO$_{2}$), which crystallizes
in the tetragonal structure of rutile-type,
is a well-known material because of its half-metallic nature
with $T_c=390$ K.\cite{Schwarz86}
The origin of the ferromagnetic (FM) and half-metallic property of CrO$_{2}$
was explained in terms of
the double-exchange model.\cite{Korotin98,Katsnelson08}
Due to the crystal field of distorted (flattened) Cr-O$_{6}$ octahedra,
Cr $t_{2g}$ states are split into lower \emph{d}$_{xy}$
and higher \emph{d}$_{xz}$/\emph{d}$_{yz}$ states.
Out of two $d$ electrons of Cr$^{4+}$, one occupies the lower \emph{d}$_{xy}$
that is localized, while the other occupies the higher
\emph{d}$_{xz}$/\emph{d}$_{yz}$ that are delocalized
near the Fermi level ($E_F$) due to the hybridization with O $p$ states.
Then the double-exchange interaction arises from the Hund coupling between
localized \emph{d}$_{xy}$ and delocalized half-filled
\emph{d}$_{xz}$/\emph{d}$_{yz}$ states, so as to produce the FM
and half-metallic properties.

In contrast to numerous reports on electronic and magnetic properties
of CrO$_{2}$, there have been relatively small number of studies on
structural and lattice dynamical properties of CrO$_{2}$.
Especially, there is no experimental or theoretical report
on the phonon dispersion curve for CrO$_{2}$,
except for a few Raman studies.\cite{Maddox06,Iliev99,Yu03}
Under the pressure, CrO$_{2}$ is known to undergo the structural transition
from the ground state tetragonal CrO$_{2}$ (t-CrO$_{2}$)
to the orthorhombic CrO$_{2}$ (o-CrO$_{2}$)
of CaCl$_{2}$-type at P = 12-14 Gpa.\cite{Maddox06,Kuznetsov06}
The question followed is whether there will an additional
structural transitions from o-CrO$_{2}$
at higher pressure. In fact, this question is not just for CrO$_{2}$
but also relevant to the structural stability issue
of transition-metal (TM) dioxides (TMO$_{2}$). Note that TMO$_{2}$s
show diverse structures depending on the TM
elements.\cite{Matteiss76,Pynn76,Eyert00,Eyert02}
Furthermore, magnetic properties of CrO$_{2}$ under the pressure
are intriguing, such as (i) whether the half-metallic nature is preserved,
and (ii) when CrO$_{2}$ becomes non-magnetic.

In this work, to investigate the pressure-induced
structural transitions of CrO$_{2}$,
we have studied phonon dispersions and total energies
of relevant CrO$_{2}$ structures as a function of pressure.
Based on the calculated phonon dispersions and the total energies,
we have found three possible structural transitions with increasing pressure.
The first transition is consistent with the known transition
from t-CrO$_{2}$ to o-CrO$_{2}$.
At this transition, FM and half-metallic properties are
preserved, in agreement with previous reports of
literature.\cite{Maddox06,Kuznetsov06}
The second transition is from o-CrO$_{2}$ to monoclinic CrO$_{2}$
(m-CrO$_{2}$) of MoO$_{2}$ type, which is nonmagnetic (NM).
The third transition is identified from m-CrO$_{2}$
to cubic CrO$_{2}$ (c-CrO$_{2}$)
of CaF$_{2}$-type. Interestingly, c-CrO$_{2}$ is a FM insulator
even at the high pressure of P $\geq$ 88.8 GPa.
Note that the second and third structural transitions are our new findings
for CrO$_{2}$ under the high pressure.

\section{Computational Details}

Band structures and phonon dispersions of CrO$_{2}$ were obtained
by employing the pseudo-potential band method and the
linear response method, respectively,
implemented in the Quantum ESPRESSO code.\cite{Giannozzi09,Comp}
The generalized gradient approximation (GGA) is utilized for
the exchange-correlation potential.
Self-consistent electron and phonon band calculations were carried out
after the full-relaxation of internal atomic positions and lattice parameters.

We have considered various structures of CrO$_{2}$.
At the ambient pressure, the stable phase is t-CrO$_{2}$
of rutile-type (\emph{P}4$_{2}$/\emph{mnm}),
in which Cr atoms are positioned at (0,0,0) and
($\frac{1}{2}$, $\frac{1}{2}$, $\frac{1}{2}$), while O atoms at
$\pm$(u, u, 0) and $\pm$($\frac{1}{2}$+u, $\frac{1}{2}-$u, $\frac{1}{2}$).
Initial lattice constants and atomic positions adopted
before the full-relaxation are
a=b=4.421${\AA}$, c=2.916${\AA}$, and u=0.3043.\cite{Xue-Wei07}
For the high pressure phase of o-CrO$_{2}$ of CaCl$_{2}$-type (\emph{Pnnm}),
we have adopted a=4.3874${\AA}$, b=4.2818${\AA}$,
c=2.8779${\AA}$, u$_{x}$ =0.299, and u$_{y}$=0.272.\cite{Maddox06}
For candidate structural phases at the higher pressure,
we considered m-CrO$_{2}$ of MoO$_{2}$ type
(\emph{P}2$_{1}$/\emph{c})\cite{Eyert00,Eyert02}
and c-CrO$_{2}$ of CaF$_{2}$-type
(\emph{Fm}$\bar{3}$\emph{m}).\cite{Srivastava08}
In the latter, a Cr atom is positioned at (0, 0, 0),
and O atoms at (0.25, 0.25, 0.25), (0.25, 0.25, 0.75).

\section{Results}

\begin{center}
\begin{figure}[t]
  \includegraphics[width=8.2 cm]{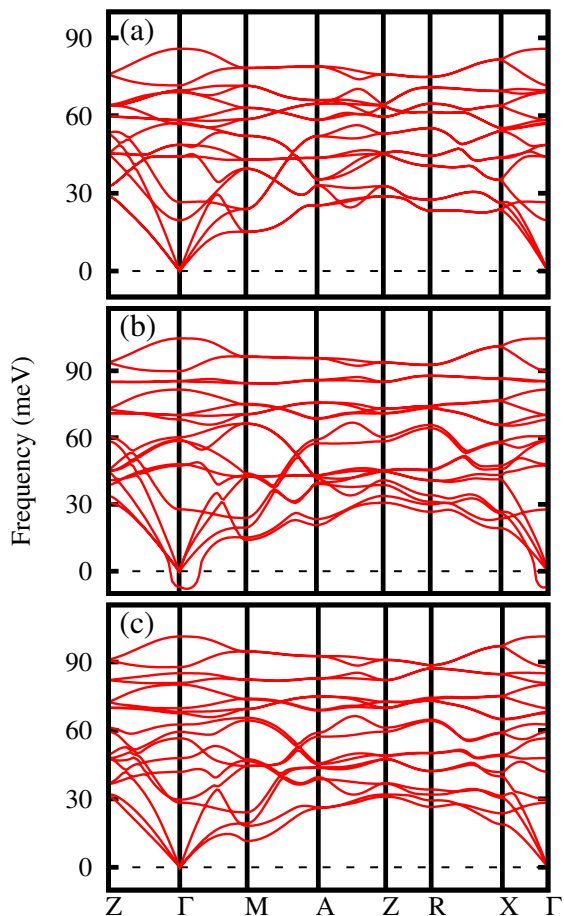}
  \caption{(Color online) The phonon dispersion curves of CrO$_{2}$.
(a) FM t-CrO$_{2}$ of rutile type at the ambient pressure.
(b) FM t-CrO$_{2}$ at P=34.0 GPa.
Notice the phonon softening at $\Gamma$,
which corresponds to the B$_{1g}$ mode.
The negative frequency here represents the imaginary part of
the phonon frequency.
(c) FM o-CrO$_{2}$ of CaCl$_{2}$ type at P=34.1 GPa.
}\label{phd_0_35}
\end{figure}
\end{center}
\begin{center}
\begin{figure}[b]
  \includegraphics[width=4 cm]{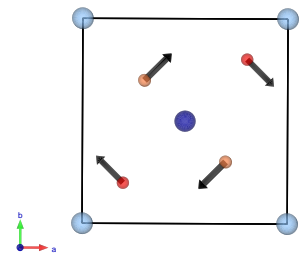}
  \caption{(Color online)
The normal mode of the B$_{1g}$ soft phonon at $\Gamma$ for P=34.0 GPa.
  The blue and light-blue circles represent Cr ions, while
  the orange and red circles represent oxygen ions.
  Blue and orange ions are located at z=$\frac{1}{2}$.
  Only the oxygen ions move in this mode.
}\label{B1g}
\end{figure}
\end{center}

Figure $\ref{phd_0_35}$ shows the phonon dispersions of t-CrO$_{2}$ and
o-CrO$_{2}$ at the ambient and high pressures.
As shown in Fig.~\ref{phd_0_35}(a), t-CrO$_{2}$ at the ambient pressure has
regular phonon dispersions, reflecting the stable phase of t-CrO$_{2}$
at the ambient pressure.
In contrast, t-CrO$_{2}$ at P=34.0 GPa in Fig.~\ref{phd_0_35}(b) has
a softening phonon mode at {\bf q}=$\Gamma$,
indicating the structural instability of t-CrO$_{2}$ at this pressure.
The softened mode corresponds to B$_{1g}$ mode that is Raman-active.
As shown in Fig.~\ref{B1g}, B$_{1g}$ mode
generates the rotating motions of oxygen ions.
The resulting lattice displacements induce the structural transformation
from t-CrO$_{2}$ of rutile type to o-CrO$_{2}$ of CaCl$_{2}$ type.
Figure $\ref{phd_0_35}$(c) provides the phonon dispersion of o-CrO$_{2}$
at P=34.1 GPa. The phonon dispersion is regular, implying that
o-CrO$_{2}$ is stable at this pressure.
Therefore, Fig.~\ref{phd_0_35} clearly demonstrates that
there is a structural transition from t-CrO$_{2}$ to o-CrO$_{2}$
at the pressure of P$\le 34$ GPa.

\begin{center}
\begin{figure}[t]
  \includegraphics[width=8.3 cm]{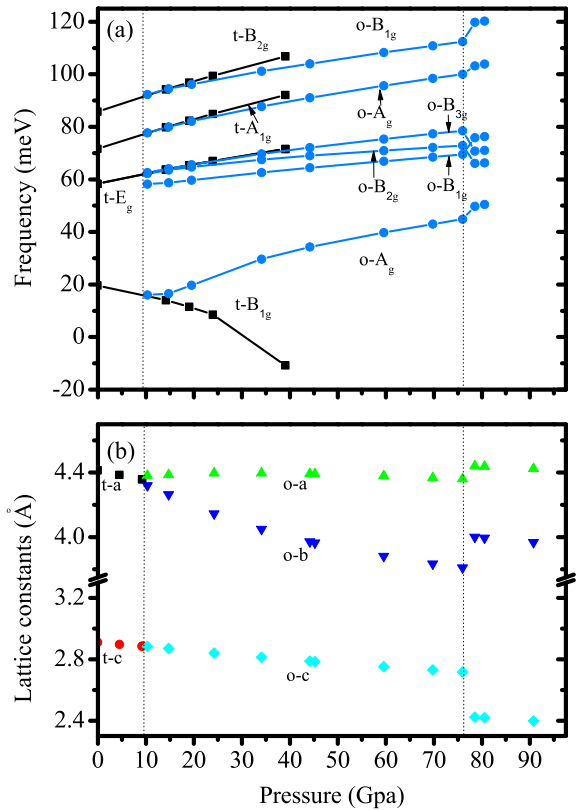}
  \caption{(Color online)
(a) Calculated Raman-active phonon frequencies of t-CrO$_{2}$
and o-CrO$_{2}$ versus pressure.
t- and o- stand for t-CrO$_{2}$ and o-CrO$_{2}$ respectively.
The lines connecting data are guide for eyes.
Two phase transitions are noticed at P=9.8 GPa and P=76.0 GPa,
which are marked by vertical lines.
(b)
Calculated equilibrium lattice constants
of t-CrO$_{2}$ and o-CrO$_{2}$ versus pressure.
}\label{lattice}
\end{figure}
\end{center}

In Fig.~\ref{lattice}(a), we plotted the calculated
Raman-active phonons of t-CrO$_{2}$ and o-CrO$_{2}$
as a function of pressure.
There are four Raman modes (B$_{1g}$, E$_{g}$, A$_{1g}$, B$_{2g}$)
for t-CrO$_{2}$, and six Raman modes
(A$_{g}$, B$_{1g}$, B$_{2g}$, B$_{3g}$, A$_{g}$, B$_{1g}$)
for o-CrO$_{2}$.\cite{Iliev99,Rosenblum97,Weber90}
Our data are consistent with experimental data
up to P$\approx 40$ Gpa.\cite{Maddox06}
With increasing the pressure,
one can clearly see the softening of B$_{1g}$ mode of t-CrO$_{2}$,
which indicates the structural instability of t-CrO$_{2}$.
One can also notice two transition points.
The first one corresponds to the transition from t-CrO$_{2}$ to o-CrO$_{2}$
at P=9.8 GPa. At this transition, CrO$_{2}$ keeps its FM and half-metallic
properties.\cite{Maddox06,Kuznetsov06,Srivastava08}.
The second one corresponds to the transition at P=76.0 GPa.
The stable structure for P$\geq 76$ has not been identified yet.

The phonon anomalies at two transition points are also revealed
in the variation of lattice constants of CrO$_{2}$ under the pressure.
In Fig.~\ref{lattice}(b), the lattice constants are plotted
as a function of pressure. One can see anomalous behaviors of
the lattice constants at the two transition points
that are coincident with those in Fig.~\ref{lattice}(a).

\begin{center}
\begin{figure}[t]
  \includegraphics[width=8.4 cm]{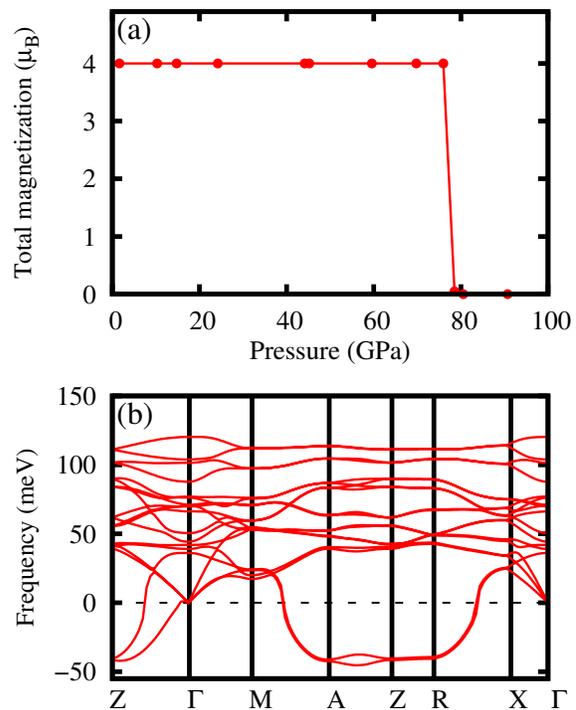}
  \caption{(Color online)
(a) Calculated magnetic moment of o-CrO$_{2}$ versus pressure.
Magnetic transition from the FM to the NM phase occurs at P=76.0 GPa.
(b) The phonon dispersion curve of NM o-CrO$_{2}$ at P=80.5 GPa.
}\label{phd_80}
\end{figure}
\end{center}

To investigate the second structural transition in more detail,
we have examined the behavior of magnetic moment.
Srivastava {\it et al.}\cite{Srivastava08} once reported that
there would be a magnetic transition in t-CrO$_{2}$
from half-metallic to NM at P$\approx 65$ GPa.
However, as discussed above,
there occurs a structural transition from t-CrO$_{2}$
to o-CrO$_{2}$ at the low pressure of about P=10 GPa.
Hence, in Fig.~\ref{phd_80}(a),
we have examined the magnetic moment behavior for o-CrO$_{2}$.
It is seen that there is a FM to NM transition at P=76.0 GPa,
which is close to the second structural transition point.

The magnetic moment of o-CrO$_{2}$ suddenly drops
at this transition point.
The half-metallic property persists up to this pressure.
This magnetic transition was also observed
by Kuznetsov \emph{et al.}\cite{Kuznetsov06},
who obtained the transition pressure of P=53 GPa
based on the pseudo-potential band method implemented in the VASP code.

It is thus tempting to conjecture that the second transition
observed in Fig.~\ref{lattice}(a)
corresponds to the magnetic transition in o-CrO$_{2}$.
However, the phonon dispersion curve in Fig.~\ref{phd_80}(b)
for NM o-CrO$_{2}$ at P=80.5 GPa shows the strong phonon
softenings, indicating that even NM o-CrO$_{2}$ is unstable
at the pressure of P$> 76.0$ GPa.
Therefore, it is not possible that FM o-CrO$_{2}$ changes
into NM o-CrO$_{2}$ with increasing the pressure.
There might be an additional
structural transition in this pressure range.
\begin{center}
\begin{figure}[b]
  \includegraphics[width=5.5 cm]{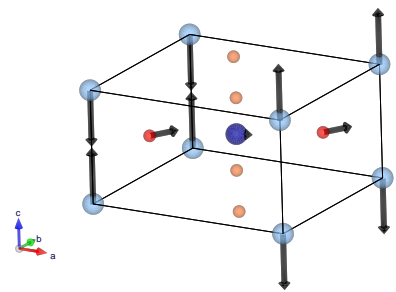}
  \caption{(Color online)
The normal mode of the softened phonon at {\bf q} = R  for P=80.5 GPa.
  The blue and light-blue circles represent Cr ions, while
  the orange and red circles represent oxygen ions.
}\label{2soften}
\end{figure}
\end{center}

Two candidate structures after the transition are
m-CrO$_{2}$ of MoO$_{2}$ type and c-CrO$_{2}$ of CaF$_{2}$ type.
The monoclinic structure of MoO$_{2}$ type is chosen
from the expectation that, with increasing pressure,
two $d$ electrons of Cr become itinerant, and
the local environment becomes similar to that of MoO$_{2}$.
In a similar system VO$_{2}$, a softening of phonon frequency
was observed at {\bf q}=R ($\frac{1}{2}$, 0, $\frac{1}{2}$),
which corresponds to the atomic movements
from the tetragonal structure of rutile-type to the
monoclinic structure.\cite{Gervais85}
Indeed, the softening mode at {\bf q} = R in Fig.~\ref{phd_80}(b) is related to
this structural transition,
because the orthorhombic structure of CaCl$_{2}$ type is nothing but
the distorted rutile-type structure.
Figure~\ref{2soften} depicts the normal mode of the softened phonon mode
at {\bf q} = R. The displacements generate Cr-Cr dimerization along the c-axis,
which is consistent with the main distortions of the transition
from the rutile-type to the monoclinic structure.
It is thus reasonable to expect the second transition to be
from o-CrO$_{2}$ of CaCl$_{2}$ type to m-CrO$_{2}$ of MoO$_{2}$ type.
Concerning another candidate, c-CrO$_{2}$ of CaF$_{2}$ type,
there were previous reports predicting the structural transitions
from CaCl$_{2}$ type to CaF$_{2}$ type structure
for CrO$_{2}$\cite{Srivastava08} and RuO$_{2}$.\cite{Haines93}
Also, CaF$_{2}$ type is a typical structure of TMO${_2}$.
For example, ZrO$_{2}$ and HfO$_{2}$, which have relatively large cations,
crystallize in CaF$_{2}$ type structure at high temperature.
\cite{Munoz-Paez94}

\begin{center}
\begin{figure}[t]
   \includegraphics[width=8.7cm]{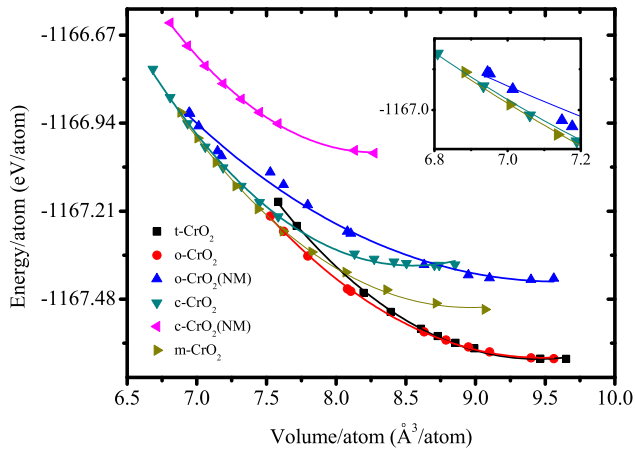}
\caption{(Color online) Total energies of various CrO$_{2}$ structures
versus volume.
Data are fitted by the Birch-Murnaghan equation of state. }\label{energy}
\end{figure}
\end{center}

To identify the additional structural transition at the higher pressure,
we have compared the total energies of candidate structures
 in Fig.~\ref{energy}.
From the total energy versus volume curves in Fig.~\ref{energy},
one can identify three structural phase transitions.
The first one is from FM t-CrO$_{2}$ to FM o-CrO$_{2}$ at
the estimated pressure of P=12.2 GPa,
which is consistent with phonon calculation in Fig.~\ref{lattice}.
For P $\geq$ 61.1 GPa, NM m-CrO$_{2}$ becomes the most stable,
which corresponds to the second transition from o-CrO$_{2}$ to m-CrO$_{2}$,
as discussed in the phonon study of Fig.~\ref{phd_80}.
For P $\geq$ 88.8 GPa, c-CrO$_{2}$, which is a FM insulator, becomes
the most stable.
The more stable FM and insulating phase than the NM metallic phase
of c-CrO$_{2}$ at this high pressure is extraordinary.
The magnetic moment of c-CrO$_{2}$ amounts to $\sim 2\mu$$_{B}$/Cr,
which is close to those in t-CrO$_{2}$ and o-CrO$_{2}$.
Note, however, that c-CrO$_{2}$ is an insulator not a half-metal.
We have confirmed this result by employing the all-electron
full potential linearized augmented plane wave method (FLAPW)
band method\cite{FLAPW}
implemented in WIEN2k package too.\cite{wien2k}
The present result is different from that
by Srivastava {\it et al.},\cite{Srivastava08}
who obtained the stable NM metallic phase of c-CrO$_{2}$ for P$>$90 GPa.
The different result is likely to come from their use of
a simple tight-binding LMTO band method.

The transition from m-CrO$_{2}$ to c-CrO$_{2}$ is thought to originate from
the increasing packing ratio.
There are six and eight oxygens around Cr in m-CrO$_{2}$
and c-CrO$_{2}$, respectively.
Haines \emph{et al.}\cite{Haines93} proposed several possible paths
of structural transition from rutile to CaF$_{2}$ type structure
in TMO$_2$.
Interestingly, one of the paths is the same
as the present structural transition path,
rutile-type (\emph{P}4$_{2}$/\emph{mnm}) $\to$
CaCl$_{2}$-type (\emph{Pnnm}) $\to$
MoO$_{2}$-type (\emph{P}2$_{1}$/\emph{c}) $\to$
CaF$_{2}$-type (\emph{Fm}$\bar{3}$\emph{m}).
But they did not take into account the magnetic state.

\section{Conclusion}

We have studied the pressure effect on the
structural properties of CrO$_{2}$ by performing the phonon dispersion
and total energy band structure calculations.
Combining two analysis methods, we have found
that there are three structural transitions with increasing pressure
up to 100 GPa.
The first one is the transition from t-CrO$_{2}$ of rutile type
to o-CrO$_{2}$ of CaCl$_{2}$ type at P$\approx$ 10 GPa
(9.8 GPa from phonon dispersion analysis, while 12.2 GPa
from total energy study).
The FM and half-metallic properties of CrO$_{2}$ are preserved
at this transition.
The second structural transition is from FM o-CrO$_{2}$ to NM m-CrO$_{2}$,
which corresponds to the lattice displacement of the phonon softening
at R in o-CrO$_{2}$. The transition pressure is P=76.0 GPa
from the phonon dispersion analysis,
whereas P=61.1 GPa from the total energy study.
The third structural transition is from NM m-CrO$_{2}$ to
FM c-CrO$_{2}$ at  P= 88.8 GPa,
which is accompanied by the metal to insulator transition.

\begin{acknowledgments}
This work was supported by the NRF (No.2009-0079947),
and the KISTI supercomputing center (No. KSC-2011-C2-36).
\end{acknowledgments}


\begin{thebibliography}{99}
\bibitem{Schwarz86}{K. Schwarz,
	J. Phys. F: Met. Phys. {\bf 16}, L211 (1986).}
\bibitem{Korotin98}{M. A. Korotin, V. I. Anisimov, D. I. Khomskii,
	and G. A. Sawatzky,
	Phys. Rev. Lett. {\bf 80}, 4305 (1998).}
\bibitem{Katsnelson08}{M. Katsnelson, V. Irkhin, L. Chioncel, A. Lichtenstein,
	and R. de Groot, Rev. Mod. Phys. {\bf 80}, 315 (2008).}
\bibitem{Maddox06}{B. R. Maddox, C. S. Yoo, D. Kasinathan, W. E. Pickett,
	and R. T. Scalettar,
	Phys. Rev. B {\bf 73}, 144111 (2006).}
\bibitem{Yu03}{T. Yu, Z. X. Shen, W. X. Sun, J. Y. Lin, and J. Ding,
	J. Phys.: Condens. Matter {\bf 15}, L213 (2003).}
\bibitem{Iliev99}{M. N. Iliev, A. P. Litvinchuk, H.-G. Lee, C. W. Chu,
	A. Barry, and J. M. D. Coey, Phys. Rev. B {\bf 60}, 33 (1999).}
\bibitem{Kuznetsov06}{A. Y. Kuznetsov, J. S. de Almeida, L. Dubrovinsky, R.
	Ahuja, S. K. Kwon, I. Kantor, A. Kantor, and N. Guignot,
	J. Appl. Phys. {\bf 99}, 053909 (2006).}
\bibitem{Matteiss76}{L. Mattheiss, Phys. Rev. B {\bf13}, 2433 (1976).}
\bibitem{Pynn76}{R. Pynn, J. Axe, and R. Thomas, Phys. Rev. B 13, 2965 (1976).}
\bibitem{Eyert00}{V. Eyert, R. Horny, K. Hock, and S. Horn,
	J. Phys.: Condens. Matter {\bf 12}, 4923 (2000).}
\bibitem{Eyert02}{V. Eyert, Ann. Phys. (Leipzig) {\bf11}, 650 (2002).}
\bibitem{Giannozzi09}{Quantum ESPRESSO (opEn-Source Package for Research
	in Electronic Structure, Simulation, and Optimization).
	P. Giannozzi et al.,
	J. Phys.: Condens. Matter {\bf 21}, 395502 (2009).
	 http://www.quantum-espresso.org.}
\bibitem{Comp}{
We have used the kinetic energy cutoff for wave functions of $\approx$ 400 eV.
We have selected {\bf k}-point samplings such that
8$\times$8$\times$8 for t-CrO$_{2}$ and o-CrO$_{2}$,
7$\times$7$\times$5 for m-CrO$_{2}$, and 12$\times$12$\times$8
for c-CrO$_{2}$ in the Monkhorst-Pack grid to set
similar {\bf k}-point density in the Brillouin zone of each structure.
}
\bibitem{Xue-Wei07}{W. Xue-Wei, N. Dong-Lin, and L. Xiao-Jun,
	Chinese Phys. Lett. {\bf 24}, 3509 (2007).}
\bibitem{Srivastava08}{ V. Srivastava, M. Rajagopalan, and S.P. Sanyal,
	The European Phys. J. B {\bf 61}, 131 (2008).}
\bibitem{Rosenblum97}{S. S. Rosenblum, W. H. Weber, and B. L. Chamberland,
	Phys. Rev. B {\bf 56}, 529 (1997).}
\bibitem{Weber90}{W. H. Weber, G. W. Graham, and J. R. McBride,
	Phys. Rev. B {\bf 42}, 10969 (1990).}
\bibitem{Gervais85}{F. Gervais and W. Kress,
	Phys. Rev. B {\bf 31}, 4809 (1985).}
\bibitem{Haines93}{J. Haines and J. M. L\'{e}ger,
	Phys. Rev. B {\bf 48}, 13344 (1993). }
\bibitem{Munoz-Paez94}{A. Mu\~{n}oz-P\'{a}ez,
	J. Chem. Edu. {\bf 71}, 381 (1994). }
\bibitem{FLAPW} B. I. Min, H. J. F. Jansen, and A. J. Freeman,
	Phys. Rev. B {\bf 33}, 6383 (1986).
\bibitem{wien2k} B. Blaha, K. Schwarz, G. K. H. Madsen, D. Kvasnicka,
	and J. Luitz,
	WIEN2K (Technische Universit\"{a}t Wien, Austria, 2001).
\end{thebibliography}
\end{document}